\documentclass{article}
\usepackage{amsmath}
\usepackage{amsfonts}
\usepackage{caption}
\usepackage{hyperref}
\usepackage{placeins}
\usepackage{graphicx} 
\usepackage{makecell}
\usepackage{iclr2025_conference,GEM_workshop_2025,times}

\title{IgCraft: A versatile sequence generation framework for antibody discovery and engineering}
\author{Matthew Greenig, Haowen Zhao, Vladimir Radenkovic, Aubin Ramon, Pietro Sormanni \\
Yusuf Hamied Department of Chemistry \\
University of Cambridge \\
Cambridge, United Kingdom, CB2 1EW \\
\texttt{\{mg989,hz362,vr375,ar2003,ps589\}@cam.ac.uk} 
}

\iclrfinalcopy

\date{January 2025}
\begin{document}

\maketitle

\begin{abstract} 
Designing antibody sequences to better resemble those observed in natural human repertoires is a key challenge in biologics development. We introduce IgCraft: a multi-purpose model for paired human antibody sequence generation, built on Bayesian Flow Networks. IgCraft presents one of the first unified generative modeling frameworks capable of addressing multiple antibody sequence design tasks with a single model, including unconditional sampling, sequence inpainting, inverse folding, and CDR motif scaffolding. Our approach achieves competitive results across the full spectrum of these tasks while constraining generation to the space of human antibody sequences, exhibiting particular strengths in CDR motif scaffolding (grafting) where we achieve state-of-the-art performance in terms of humanness and preservation of structural properties.  By integrating previously separate tasks into a single scalable generative model, IgCraft provides a versatile platform for sampling human antibody sequences under a variety of contexts relevant to antibody discovery and engineering. Model code and weights are publicly available at \url{https://github.com/mgreenig/IgCraft}.
\end{abstract}

\section{Introduction}

Monoclonal antibodies are an important class of therapies that comprise an increasingly large share of the global pharmaceutical market \citep{Ecker2015}. Key to the success of these molecules as therapeutics lies not only in their ability to selectively bind their target with high affinity, but also in their favorable \textit{developability}, a property that broadly describes the suitability of a functional compound to become a viable drug, often a function of immunogenicity, solubility, and a number of other factors. Conventional antibody discovery typically relies on either animal immunization \citep{Lee2014} or high-throughput screening of large sequence libraries \citep{Bradbury2011} to isolate potential candidates. While \textit{in vitro} screening methods are faster, cheaper, and have ethical advantages compared to immunization, naturally-derived antibodies tend to exhibit better developability properties, including favorable pharmacokinetics, high specificity, and low immunogenicity \citep{Jain2017}. It is therefore no surprise that machine learning models trained on large natural sequence databases have been successfully developed for many specific tasks in antibody developability engineering, including de-novo sequence generation \citep{Turnbull2024}, conditional design of subsequences \citep{Olsen2024}, structure-conditioned sequence design \citep{Dreyer2023, Hoie2024}, and grafting of mouse complementarity-determining regions (CDRs) into suitable human frameworks \citep{Ma2024}. However, the deployment of multiple task-specific
models introduces additional complexity into computational workflows and can
hinder their scalability and broader adoption. Here, we present a unified generative modeling framework that addresses multiple antibody engineering challenges simultaneously, achieving performance in most cases that is competitive with or surpasses state-of-the-art task-specific models. 

IgCraft is a generative model that uses a Bayesian Flow Network (BFN) \citep{Graves2023, Atkinson2025} to sample paired human antibody sequences under a variety of contexts relevant to therapeutic antibody development. Similar to diffusion models, BFNs perform a denoising process jointly across all sequence positions, decoupling the number of generation steps from sequence length and enforcing no particular generation order amongst the tokens (unlike autoregressive models). Crucially, the BFN's lack of a specific generation order enables flexible sequence inpainting (conditional generation of subsequences) from a single model trained only to generate full-length sequences, as was demonstrated in ProtBFN \citep{Atkinson2025} with a sequential monte carlo approach. The ability to perform inpainting for arbitrary sequence regions is especially attractive given the multitude of design scenarios encountered in therapeutic antibody development, including developability optimisation, humanisation, and affinity maturation. As a model that can efficiently perform both unconditional and conditional sequence sampling with the flexibility to incorporate structural data (when available), IgCraft offers substantial practical advantages as a single tool that can be used for a variety of tasks in real-world antibody engineering workflows.

\section{Methods}

\subsection{Bayesian flow networks}

A discrete-variable Bayesian Flow Network (BFN) is a generative model over discrete tokens $\{ x_i \in V \}_{i=1}^D$ for some vocabulary $V$ of size $K$. BFNs are conceptually similar to diffusion models \citep{Ho2020, Austin2021}, but instead of modeling a discrete denoising process over tokens themselves, BFNs model a continuous denoising process over vectors of logits for different token categories $\{ {\bf z}_i \in \mathbb{R}^K \}_{i=1}^D$. Specifically, the BFN generative process for any given token can be formulated in terms of an SDE in logit space \citep{Xue2024}:

\begin{equation}
    d{\bf z} = \alpha(t) \left [ K\hat{e}({\bf z}(t), t) - {\bf 1} \right ]dt + \sqrt{K\alpha(t)}d{\bf w}
    \label{bfn_inference_sde}
\end{equation}

Where $\hat{e}({\bf z}(t), t) \in \Delta^K$ is a predicted vector of probabilities over token categories and $\alpha(t) \in \mathbb{R}^+$ is an accuracy schedule, playing a similar role to the variance schedule in a diffusion model. In practice, $\hat{e}$ is a neural network that is given the current logits for all sequence positions as input and uses learned relationships between these noisy variables to make a prediction for each token's value ${\bf x}$.  We use $\alpha(t) = 2t$ for all experiments as in the original BFN paper \citep{Graves2023}. We also introduce a temperature parameter that scales the network's output logits before they are converted into probabilities via softmax and use $T = 1.05$ for unconditional sampling and $T = 0.1$ for all other (conditional) sampling tasks. To solve the SDE in \eqref{bfn_inference_sde}, we implement a second-order solver similar to that proposed by \cite{Xue2024} and perform sampling in 20 steps. For conditional sampling, we use the particle filtering method outlined in ProtBFN \citep{Atkinson2025} with 32 particles. More details on IgCraft's sampling methodology can be found in \autoref{appendix}.

\subsection{Network architecture}

Since the BFN is agnostic to choice of network architecture, we introduce a two-track transformer configuration designed to model paired antibody sequences (\autoref{figure:architecture}). To process sequence tokens within each antibody chain individually, the architecture makes use of standard transformer blocks with gated self-attention \citep{Chai2020}, rotary positional embeddings \citep{Su2021}, pre-layer normalization \citep{Xiong2020}, and SwiGLU transition layers \citep{Shazeer2020}, with one transformer stack allocated to process VH sequence tokens and the other to process VL sequence tokens. After each transformer block, token embeddings for both the VH and VL chains are fed into an \textit{interaction block} that uses gated cross-attention, adaptive layer normalization (AdaLN) \citep{Xu2019}, and conditional SwiGLU transition layers \citep{Abramson2024} to integrate information from tokens in the other chain. For the AdaLN and conditional SwiGLU layers, the mean token embedding from the other chain's sequence is used as conditioning data for each token's update. The output of the interaction block is projected via a sigmoid-gated linear unit and fed into a residual connection with the transformer stack's embeddings after processing. To encode structural information, we use the geometric multi-head attention architecture from ESM3 \citep{Hayes2025} with separate embedding layers for VH, VL, and epitope residues. All together IgCraft contains approximately 300M trainable parameters.

\subsection{Datasets and training regime}

\begin{figure}[h]
\begin{center}
\includegraphics[scale=0.425]{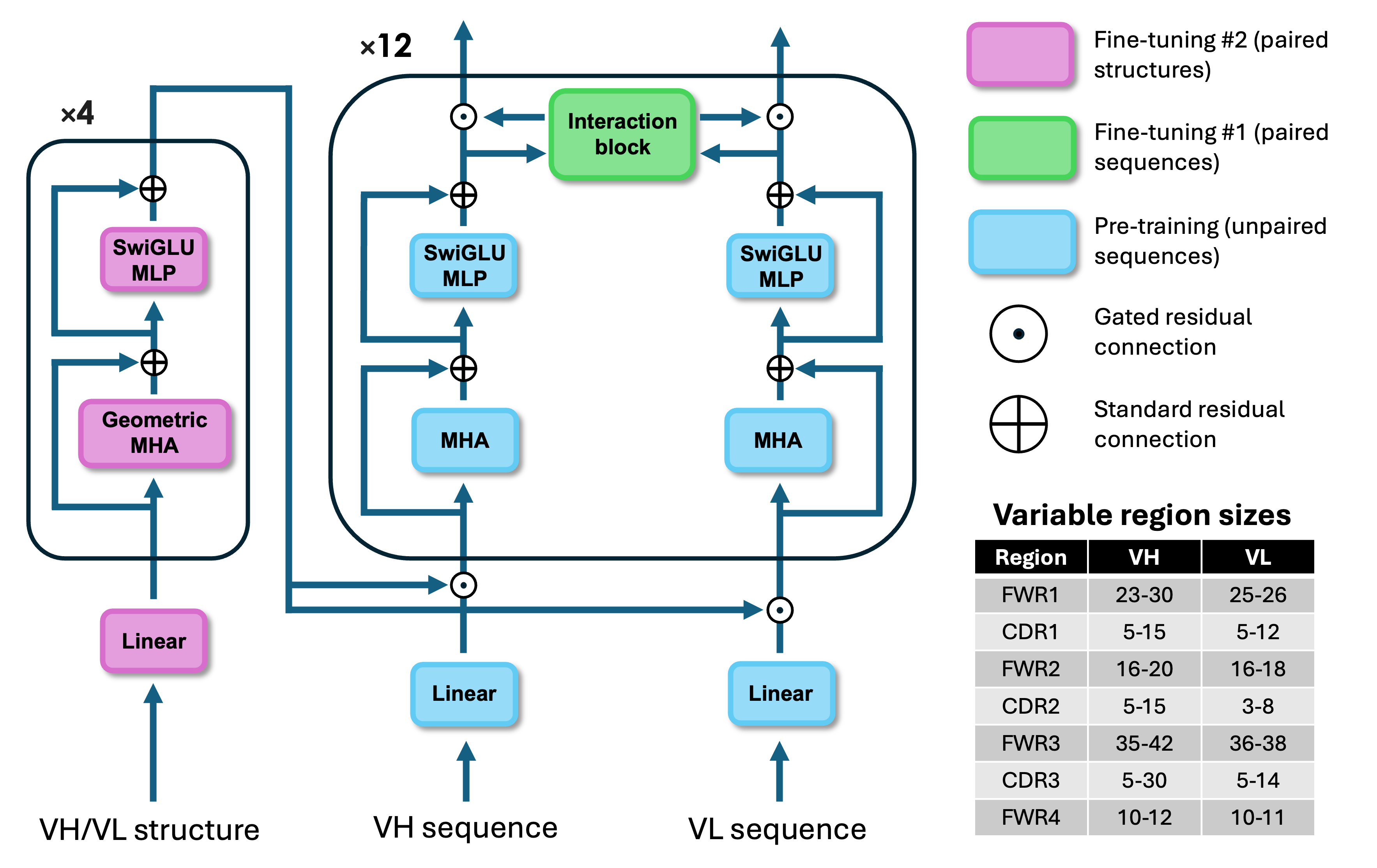}
\end{center}
\captionsetup{font=footnotesize}
\caption{IgCraft's two-track transformer architecture. Layers are color-coded by the stage of training during which they are updated. The main backbone (blue, green) receives noisy logits for the VH/VL sequences as input and outputs predicted probabilities for the amino acid identity at each position. Shown in the bottom right corner are the minimum/maximum lengths (both sides inclusive) per variable domain region of each antibody chain type. MHA: Multi-head attention; MLP: Multi-layer perceptron; SwiGLU: Swish-gated linear unit.}
\label{figure:architecture}
\end{figure}

 To obtain variable region annotations (FWR1, CDR1, etc. according to IMGT definition), we merged the set of paired and unpaired antibody sequences from \cite{Turnbull2024} with the Observed Antibody Space database \citep{Olsen2021}. We enforced minimum and maximum length cutoffs per-region (\autoref{figure:architecture}, bottom right), with values determined by qualitative analysis of the distribution of region lengths in OAS. To enable the model to generate sequences of varying length under different conditional design scenarios, we perform padding within each variable domain region of each input sequence, right-padding to a maximum length per-region. The model is then trained to generate pad tokens (as well as amino acid tokens) to control the length of generated sequences. Using the same train/test/validation splits as in \cite{Turnbull2024}, our filtering process yielded training sets of 118M unpaired VH sequences, 135M unpaired VL sequences, and 1.5M paired VH/VL sequences. For testing we use a similarly filtered subset of the paired test sequences from \cite{Turnbull2024}, which yielded a test set of 63,705 paired sequences. To obtain structural data for fine-tuning, we clustered the training set of 1.5M paired VH/VL sequences (concatenated) at 40\% minimum sequence identity using MMSeqs2 \citep{Steinegger2017}, folded each cluster's representative sequence using ABodyBuilder3-LM \citep{Kenlay2024}, and removed structures with mean H-CDR3 pLDDT \textless70, producing a set of approximately 30,000 predicted structures. We merged these predicted structures with a curated set of approximately 2,800 non-redundant human paired VH/VL structures extracted from SAbDab \citep{Schneider2021} (details in \autoref{appendix}). For bound SAbDab structures, a maximum of 128 non-antibody (target) residues with the lowest C$\alpha$ atom distance to the antibody are included in each structure. For inverse folding tasks we use the subset of chains from our set of 2,800 unique human paired antibody structures whose PDB IDs appear in the test set but not the training/validation sets of AbMPNN \citep{Dreyer2023}, leaving 98 structures for testing. CDR grafting was tested on a holdout set of 27 paired mouse antibody structures deposited in the PDB from February 2024 onwards.

Training IgCraft consists of three stages. First, each chain's transformer stack (\autoref{figure:architecture}, blue) is pre-trained on unpaired sequences. Then, the model is fine-tuned on paired sequences, for which the pre-trained weights for both transformer stacks are loaded into the network and the interaction blocks (\autoref{figure:architecture}, green) are randomly initialized. However, we initialize the bias term in the output gate of each interaction block to a value of $-5.0$, negating its contribution to the token embeddings at the start of fine-tuning and effectively initializing the model as two unpaired sequence models that do not communicate. All weights are updated during this stage of fine-tuning, including the weights initialized from pre-training. Finally, the model is fine-tuned using paired antibody structures as conditioning information for the model's sequence predictions, initializing the output gate of the structure encoder to $-5.0$. During structure-fine-tuning, the framework regions of input structures are masked stochastically in 50\% of training examples to train the model to perform CDR-conditional framework generation. In this stage, only the weights in the structure encoder are updated to ensure that the main trunk retains its capabilities as a pure sequence generative model. This approach is conceptually similar to the approach proposed by \cite{Zheng2023} for performing inverse folding by augmenting a protein language model with a lightweight structural adapter.

\section{Results}

\subsection{Unconditional sequence generation}

We first evaluated the model's ability to sample human-like paired antibody sequences. Specifically, we generated 2000 paired sequences unconditionally using both IgCraft and p-IgGen \citep{Turnbull2024} and calculated statistics between each sample and the test set of native paired sequences. For IgCraft sampling we used a temperature of 1.05 and for p-IgGen all sampling defaults were used, including removing the bottom 5\% of sequences (ranked by perplexity). In addition, a set of 2,000 real paired antibody sequences were held out from the test set and treated as samples to calculate reference statistics. Results for IgCraft, p-IgGen, and the reference set are shown in \autoref{table:unconditional}.

\begin{table}[t]
\captionsetup{font=footnotesize}
\caption{Unconditional paired human antibody sampling results. To measure novelty, we calculate each sequence's minimum edit distance to any sequence in the test set, while for diversity, we calculate each sequence's minimum edit distance to any sequence in its own set of samples. Displayed are the means of these values over all samples in each set, shown separately for the VH and VL chains. As in p-IgGen \citep{Turnbull2024}, VH/VL pairing compatibility is estimated via the pearson correlation between germline sequence identities of the heavy and light chains. }
\begin{center}
\begin{tabular}{|c|c|c|c|}
\hline 
{\bf Samples} & {\bf Novelty (VH / VL)} & {\bf Diversity (VH / VL)} & {\bf VH/VL mut. corr.} 
\\ \hline 
IgCraft & 10.5 / 2.7 & 13.0 / 3.4 & 0.49 \\
Reference & 10.2 / 2.7 & 12.8 / 3.4 & 0.55  \\
p-IgGen & 10.4 / 3.1 & 12.7 / 3.7 & 0.51 \\
\hline
\end{tabular}
\label{table:unconditional}
\end{center}
\end{table}

\subsection{Sequence inpainting}

To evaluate the model's conditional sampling capabilities we performed sequence inpainting on the same 2000 held-out test sequences from paired OAS. For each of these sequences, we masked-out and inpainted each variable region individually as well masking and inpainting multiple regions jointly. Joint inpainting was performed for CDRs and framework regions separately, both on the heavy and light chains individually (where the entire other chain was provided as conditioning information) and on both chains jointly. Sampling was performed using IgCraft with a temperature of 0.1, with AbLang2 \citep{Olsen2024} and ESM3 \citep{Hayes2025} as benchmarks. We used default sampling parameters for AbLang2 and sampled in 10 steps with temperature 0.1 for ESM3. For IgCraft, if the generated sequence was of the incorrect length, we performed a pairwise alignment with the ground truth and calculated amino acid recovery as the percentage of matching positions normalized by the total length of the alignment, including gaps. The mean AARs for single-CDR inpainting as well as VH/VL joint inpainting for all CDRs and all framework regions are shown in \autoref{table:inpainting}, with additional data in appendix \ref{appendix}.

\begin{table}[t]
\captionsetup{font=footnotesize}
\caption{Mean amino acid recovery (AAR) for sequence inpainting on the holdout set of 2000 paired sequences. Shown are the AARs for inpainting each of the heavy chain CDRs individually as well as jointly inpainting all CDRs and all framework regions on both chains. Since ESM3-open only accepts single-chain inputs, to model paired antibody sequences we concatenate the heavy and light chains with the commonly-used (G$_4$S)$_3$ linker sequence for single-chain paired antibodies \citep{Huston1988}. We were not able to determine if this test set overlaps with the training sets of AbLang2 and ESM3.}
\begin{center} 
\resizebox{\textwidth}{!}{
\begin{tabular}{|c|c|c|c|c|c|c|c|c|c|}
\hline
{\bf Samples} & {\bf H-CDR1 (\%)} & {\bf H-CDR2 (\%)} & {\bf H-CDR3 (\%)} &  {\bf All CDRs (\%)} & {\bf All FWRs (\%)} 
\\ \hline 
IgCraft & 91.6 & 89.4 & 41.1 & 74.2 & {\bf 96.7} \\
AbLang2 & {\bf 92.0} & {\bf 90.4} & {\bf 45.5} & {\bf 76.3} & 83.3 \\
ESM3-open & 77.7 & 58.4 & 34.1 & 54.1 & 13.0 \\
\hline
\end{tabular}
}
\label{table:inpainting}
\end{center}
\end{table}

\subsection{Inverse folding and structure-guided sequence design}

To assess the model's ability to conditionally generate antibody sequences in the presence of structural information, we sampled a single sequence for each of the 98 test structures using IgCraft with a temperature of 0.1, providing the antibody and target backbone structures as input to the model's structure encoder but exposing no sequence information from the antibody (target sequence identities are used as input features). For comparison, we also performed inverse folding (also using temperature 0.1) using ProteinMPNN \citep{Dauparas2022}, AbMPNN \citep{Dreyer2023}, and Antifold \citep{Hoie2024}. While the target was cropped to 128 residues when sampling with IgCraft, the full target chains were provided as input to all other inverse folding methods. To measure performance, we calculate the amino acid recovery for each variable domain region and estimate two properties related to the developability potential of the generated sequences: humanness and solubility. Humanness scores for the generated sequences are obtained using AbNatiV \citep{Ramon2024} while solubility is scored with CamSol \citep{Sormanni2015, Rosace2023}, both of which have been supported by experimental validation. Results are shown in \autoref{table:inverse_folding}.

\begin{table}[t]
\captionsetup{font=footnotesize}
\caption{Mean amino acid recovery (AAR) and developability metrics for sequences generated by inverse folding models on a holdout set of 98 paired human antibody structures. Statistics are calculated using a single sample from each method with a temperature of 0.1. For the framework (FWR) AAR statistics we report the mean AAR over all VH/VL framework positions. To measure developability potential we estimate humanness and solubility using the AbNatiV \citep{Ramon2024} and CamSol \citep{Sormanni2015} scores respectively and report the mean scores for the VH and VL chains over all 98 sequences. We also calculated these scores for the 98 ground-truth test set antibodies and obtained VH / VL scores of 0.87 / 0.93 for AbNatiV and 0.05 / 0.45 for CamSol, demonstrating that IgCraft was the only method to improve both metrics for both chains.}
\begin{center}
\resizebox{\textwidth}{!}{
\begin{tabular}{|c|c|c|c|c|c|c|}
\hline 
{\bf Samples} & {\bf H-CDR1 (\%)} & {\bf H-CDR2 (\%)} & {\bf H-CDR3 (\%)} &  {\bf FWR (\%)} & {\bf \makecell{Humanness \\ (VH / VL)}} & {\bf \makecell{Solubility \\ (VH / VL)}} 
\\ \hline 
IgCraft & 73.5 & 66.7 & 45.0 & 91.2 & {\bf 0.91} / {\bf 0.97} & {\bf 0.08} / {\bf 0.49} \\
Antifold & {\bf 77.1} & {\bf 75.3} & {\bf 58.1} & {\bf 92.7} & 0.89 / 0.95 & 0.01 / 0.47 \\
AbMPNN & 74.0 & 65.0 & 55.6 & 87.8 & 0.85 / 0.90 & 0.02 / 0.42 \\
ProteinMPNN & 48.9 & 46.4 & 31.9 & 58.9 & 0.51 / 0.51 & -0.12 / 0.11 \\
\hline
\end{tabular}
}
\label{table:inverse_folding}
\end{center}
\end{table}

\subsection{CDR grafting and humanisation}

Scaffolding ("grafting") of CDRs from mouse or other non-human antibodies into human antibody framework regions is a key task in antibody engineering \citep{Jones1986, Kim2012}, since non-human antibodies are often easier to obtain but can induce pathological anti-drug immune responses when administered as therapeutics \citep{Khazaeli1994}. Traditional CDR grafting workflows typically rely on sequence homology searches to select an existing human framework similar to the input antibody \citep{Kim2012}. However, these approaches are inherently limited by the availability of similar framework sequences and often substantially decrease binding affinity by altering key interactions between the CDR and framework regions \citep{Pavlinkova2001}. HuDiff was recently proposed as an ML-driven solution for this task \citep{Ma2024}, which uses a discrete diffusion model to conditionally generate framework sequences given a set of input CDR sequences, and is fine-tuned specifically on mouse antibodies to sample mutations that improve the AbNatiV humanness score \citep{Ramon2024} of generated sequences. The authors include impressive experimental evidence demonstrating that the binding properties of an existing high-affinity mouse antibody (\textless1nM kD) were largely preserved after humanisation. However, HuDiff provides no mechanism for integrating structural information for the input CDRs.

Given IgCraft's strong performance in generating human framework sequences and its capacity to condition on structural inputs, we sought to evaluate the model's ability to generate human framework sequences for scaffolding mouse CDRs. We performed conditional sampling with IgCraft using 27 mouse antibodies from SAbDab as input and generating framework sequences for both chains, providing the CDR structures as conditioning information as well as the CDR sequences $\pm$2 residues on each side. For each set of mouse CDRs, we also sampled a single paired sequence from HuDiff \citep{Ma2024} as a benchmark. The effectiveness of CDR grafting is measured in two ways: first, the extent to which the humanness of the antibody is improved after grafting (compared to the parental antibody), and second, the extent to which the binding properties of the antibody are preserved. We evaluated humanness using AbNatiV \citep{Ramon2024}, an autoencoder-based deep learning method, and OASis \citep{Prihoda2022}, an approach that estimates humanness using 9-mer peptide frequences in OAS. To test how well both grafting methods maintained the structure and binding properties of the wild-type CDRs, we applied AlphaFold3 \citep{Abramson2024} to fold each generated sequence (and its target, if applicable) with 10 seeds, as well as the 27 original mouse sequences for comparison. For each structure prediction we measure the RMSD of the H-CDR3 loop (superimposing only on the framework region), and for bound antibodies, determine whether the antibody was docked to the correct epitope on the target protein using the widely-applied DockQ score threshold of 0.23 \citep{Mirabello2024, Abramson2024}. Results are shown in \autoref{table:grafting}. We also include an ablation study in \autoref{appendix} in which CDR grafting was performed using IgCraft without structural information, demonstrating that providing CDR structures as input significantly improves the model's ability to propose framework sequences that preserve the structural features of the input CDRs. 

\begin{table}[t]
\captionsetup{font=footnotesize}
\caption{Evaluation of mouse CDR grafting on a test set of 27 paired mouse antibody structures from SAbDab (20 bound, 7 unbound). We report the mean sequence identity (\%) between the grafted and original mouse sequences, the mean OASis humanness score \citep{Prihoda2022}, the mean AbNatiV humanness scores per-chain \citep{Ramon2024}, the mean H-CDR3 C$\alpha$ RMSD between the AlphaFold3 predictions and each corresponding ground-truth PDB, and the fraction of the sequences for the 20 bound structures correctly docked by AF3 (DockQ \textgreater0.23). The reference samples refer to the original 27 mouse sequences. We note that HuDiff is specifically fine-tuned to maximize AbNatiV score, while IgCraft is not.}
\begin{center}
\resizebox{\textwidth}{!}{
\begin{tabular}{|c|c|c|c|c|c|}
\hline 
{\bf Samples} & {\bf \makecell{\% Seq. id. \\ (VH / VL)}} & {\bf \makecell{Humanness \\ (OASis)}} & {\bf \makecell{Humanness \\ (AbNatiV, VH / VL)}} & {\bf H-CDR3 RMSD (\AA)} & {\bf DockQ \textgreater0.23} 
\\ \hline 
IgCraft & 77.6 / 77.5 & \textbf{77.9} & \textbf{0.88} / \textbf{0.90} & 2.04 & \textbf{10/20} \\
HuDiff & 81.4 / 80.3 & 74.6 & 0.87 / 0.82 & \textbf{1.99} & 9/20 \\
Reference & 100.0 / 100.0 & 47.3 & 0.68 / 0.64 & 1.87 & 11/20 \\
\hline
\end{tabular}
}
\label{table:grafting}
\end{center}
\end{table}

\section{Conclusion}

This work presents, to the best of our knowledge, the first generative model for paired antibody sequences that can natively perform both unconditional and conditional sequence generation and flexibly condition on structural information. We demonstrate that IgCraft's unconditionally generated sequences recapitulate patterns of variation observed in natural human antibody repertoires (\autoref{table:unconditional}), and further show that inference-time conditional sampling can be used to achieve competitive sequence inpainting results, with IgCraft exhibiting state-of-the-art performance in particular for inpainting framework regions conditional on CDRs (\autoref{table:inpainting}). In inverse folding, IgCraft achieves amino acid recovery rates competitive with leading approaches, with performance on H-CDR3 being notably lower than state-of-the-art antibody-specific tools but superior to ProteinMPNN (\autoref{table:inverse_folding}). Importantly, IgCraft's generated sequences in inverse folding attain better humanness and solubility profiles than competing methods and demonstrate improvement over the wild-type sequences on all fronts, highlighting the tool's potential to perform structure-guided sequence design in the context of antibody developability optimisation. Finally, we demonstrate using a test set of mouse antibody structures that IgCraft's conditional framework generation is capable of grafting mouse CDRs into human antibody framework regions to increase humanness while maintaining functionality (\autoref{table:grafting}). Compared to another leading ML-based grafting approach \citep{Ma2024}, IgCraft achieves better humanisation while achieving equal or better preservation of the functional features of the parental antibody (as assessed by AlphaFold3 structure prediction). We hope to explore in future work how the aggressiveness of the humanisation strategy (and its structural properties) can be controlled by modulating sampling parameters or providing additional conditioning information. All in all, we present promising initial results indicating that a wide variety of antibody sequence generation tasks can be accomplished using a unified, scalable model architecture.

\newpage

\section{Acknowledgements} 
P.S. is a Royal Society University Research Fellow (grant no. URF\textbackslash R1\textbackslash 201461). We acknowledge funding from UK Research and Innovation (UKRI) Engineering and Physical Sciences Research Council (grant no. EP/X024733/1, an ERC starting grant to P.S. underwritten by UKRI). M.G. is a Yusuf Hamied Graduate Scholar. M.G. thanks Timothy Atkinson, Tom Barrett, Alex Graves, and the rest of the team at InstaDeep for helpful discussions and training provided during an internship. The research presented here, however, was conducted independently after the internship, relying exclusively on publicly available or independently developed methods, code, and datasets. 

\bibliography{bibliography}
\bibliographystyle{iclr2025_conference}

\newpage 

\appendix
\section{Appendix}
\label{appendix}

\subsection{Generative model details}

\subsubsection{Background on Bayesian Flow Networks}

For discrete tokens $({\bf x}_1, ..., {\bf x}_D)$, a Bayesian Flow Network attempts to approximate the following SDE over logits for each token $({\bf z}_1, ..., {\bf z}_D)$ \citep{Xue2024}:

\begin{equation}
    d{\bf z}_i = \alpha(t) \left [ Ke({\bf x}_i) - {\bf 1} \right ]dt + \sqrt{K\alpha(t)}d{\bf w}
    \label{bfn_true_sde}
\end{equation}

 Where $e({\bf x}_i) \in \mathbb{R}^K$ is a one-hot encoding of the token's value and $\theta_i = \text{softmax}({\bf z}_i)$ gives a probability distribution over token categories. From the boundary condition ${\bf z}(0) = {\bf 0}$ (uniform probabilities over token categories at the start of generation), this SDE transforms an uninformed "prior" into a distribution that becomes progressively more concentrated around the token's true value as $t \rightarrow 1$. We can obtain the conditional distribution $p({\bf z}(t) | {\bf x}, t)$ in closed form as:

\begin{gather*}
    \beta(t) = \int_0^t{\alpha(s) ds} \\
    \stepcounter{equation} {\bf z}(t) \sim \mathcal{N} \left (
        \beta(t)[Ke({\bf x}) - {\bf 1}], 
        K\beta(t){\bf I}
    \right ) \tag{\theequation} 
    \label{flow_distribution}
\end{gather*}

A single training step of the BFN is performed by sampling $t$ uniformly, sampling ${\bf z}(t) \sim p({\bf z}(t) | {\bf x}, t)$ \eqref{flow_distribution} for each token in the input, and performing a gradient step on the mean-squared error between the predicted probabilities $\hat{e}({\bf z}(t), t)$ and the ground-truth one-hot encoding for each token, i.e.:

\begin{equation}
    \mathcal{L}({\bf x}) = \mathbb{E}_{t \sim U(0, 1), {\bf z}(t) \sim p({\bf z} | {\bf x}, t)}{
    \left [ \frac{\alpha(t)}{2} \lVert \hat{e}({\bf z}(t), t) - e({\bf x}) \rVert^2 \right ]
    }
    \label{bfn_loss}
\end{equation}

For a detailed derivation of the loss in \eqref{bfn_loss} as a variational lower bound on the model likelihood for a given observation, readers should consult the original BFN work \citep{Graves2023}.

\subsubsection{BFN sampling}

Sampling from a trained BFN involves solving the SDE in \eqref{bfn_inference_sde}. The exact solution over some interval $[t_{i-1}, t_i] \subseteq [0, 1]$ is:

\begin{equation}
    {\bf z}(t_i) = {\bf z}(t_{i-1}) + K\int_{t_{i-1}}^{t_i}
    {
    \alpha(s)
    \left [ 
        \hat{e}({\bf z}(s), s)
        - \frac{1}{K} 
    \right ]
    ds
    }
    + \sqrt{K(\beta(t_i) - \beta(t_{i-1})}{\bf u}
    \label{exact_bfn_sde}
\end{equation}

With ${\bf u} \sim \mathcal{N}({\bf 0}, {\bf I})$. In their original work, \cite{Graves2023} proposed a first-order solver:

\begin{equation}
    {\bf z}(t_i) = {\bf z}(t_{i-1}) + K(\beta(t_i) - \beta(t_{i-1}))
    \left [ 
        \hat{e}({\bf z}(t_{i-1}), t_{i-1})
        - \frac{1}{K} 
    \right ]
    + \sqrt{K(\beta(t_i) - \beta(t_{i-1})}{\bf u}
\end{equation}

In subsequent work, \cite{Xue2024} implemented a second-order solver for the original discrete-variable BFN accuracy schedule $\alpha(t) = 2t\beta_1$ from \cite{Graves2023}, with $\beta_1$ as a hyperparameter. Here, we derive a simpler and more flexible form of their solver with a straightforward approximation. Using the shorthand $\hat{e}({\bf z}(t_{i}), t_i) := \hat{e}_i$, we start with the same second-order approximation to $\int_{t_{i-1}}^{t_i}{
    \alpha(s)
    \left [ 
        \hat{e}({\bf z}(s), s)
        - \frac{1}{K} 
    \right ]
    ds
}$ using finite differences:

\begin{align}
    \int_{t_{i-1}}^{t_i}{
    \alpha(s)
    \left [ 
        \hat{e}({\bf z}(s), s)
        - \frac{1}{K} 
    \right ]
    ds
    } \approx 
    \int_{t_{i-1}}^{t_i}{
    \alpha(s)
    \left [ 
        \hat{e}_{i-1} 
        - \frac{1}{K} 
        + \frac{\hat{e}_{i-1} - \hat{e}_{i-2}}{t_{i} - t_{i-1}}(s - t_{i-1})
    \right ]
    ds
    } \\ 
    = \left [ 
        \hat{e}_{i-1}
        - \frac{1}{K} 
    \right ] (\beta(t_i) - \beta(t_{i-1})) + 
    \frac{\hat{e}_{i-1} - \hat{e}_{i-2}}{t_{i} - t_{i-1}}\int_{t_{i-1}}^t{\alpha(s) (s - t_{i-1}) ds} 
    \label{second_order_term}
\end{align}

Integrating by parts the second term in \eqref{second_order_term} gives:

\begin{align}
    \int_{t_{i-1}}^{t_i}{\alpha(s) (s - t_{i-1}) ds} = \beta(t_i)(t_i - t_{i-1}) - \int_{t_{i-1}}^{t_i}{\beta(s) ds} \\
    \approx \beta(t_i)(t_i - t_{i-1}) - 
    \underbrace{\frac{(\beta(t_i) + \beta(t_{i-1}))(t_i - t_{i-1})}{2}}_{\text{Trapezoid approximation}} \\
    = \frac{(\beta(t_i) - \beta(t_{i-1}))(t_i - t_{i-1})}{2}
\end{align}

The trapezoid approximation is expected to be highly accurate since $\beta(t)$ is monotonically increasing in time. This slight modification not only simplifies the final form of the solver; it also allows for arbitrary accuracy schedules to be used in place of $\alpha(t) = 2t\beta_1$ since it only requires point evaluations of $\beta(t)$. Other work has investigated new accuracy schedules for the BFN \citep{Tao2025}, and the original BFN developers noted that $\alpha(t) = 2t\beta_1$ was chosen primarily as a heuristic, with further investigations left to future work \citep{Graves2023}.

Putting it all together and simplifying, our second-order BFN solver is:

\begin{equation}
    {\bf z}(t_i) = {\bf z}(t_{i-1}) + K(\beta(t_i) - \beta(t_{i-1})) 
    \left [ 
        \frac{
            3\hat{e}_{i-1} - 
            \hat{e}_{i-2}
        }{2}
        - \frac{1}{K} 
    \right ]
    + \sqrt{K(\beta(t_i) - \beta(t_{i-1}))} {\bf u}
\end{equation}

\subsubsection{Conditional sampling}

For conditional sampling tasks with IgCraft we implement the sequential monte carlo (SMC) framework from ProtBFN \citep{Trippe2022, Atkinson2025}. In general, SMC methods attempt to approximate a posterior distribution over the data ${\bf x}$ given some conditioning information ${\bf y}$:

\begin{equation}
    p({\bf x} | {\bf y}) \propto p({\bf y} | {\bf x})p({\bf x})
\end{equation}

In \textit{sequential importance resampling}, at each step of the sampling trajectory, proposal samples are first drawn from the unconditional model $p({\bf x})$ and then re-sampled with replacement under some appropriately normalized likelihood function $p({\bf y} | {\bf x})$. The trained BFN - via the SDE in \eqref{bfn_inference_sde} - provides access to the unconditional distribution over token logits $p({\bf z}(t))$. The problem of sequence inpainting considers the task of conditioning generation on a subset of sequence positions $M \subset \{1, 2, ..., D\}$ with corresponding tokens ${\bf x}_M := \{{\bf x}_i\}_{i \in M}$. Denoting samples of the proposal distribution as $\{ {\bf z}_p(t) \}_{p=1}^Q$, ProtBFN \citep{Atkinson2025} uses the following form for the conditional likelihood:

\begin{align}
    w_p = \sum_{i \in M}{
    -\lVert 
        \hat{e}({\bf z}_p(t), t)_i - e({\bf x}_i)
    \rVert^2
    } \\
    \label{particle_weights}
    p({\bf x}_M | {\bf z}_p(t)) = \frac{
        \exp{(w_p)}}
    {\sum_{q = 1}^Q{\exp{(w_q)}}}
\end{align}

Where $Q$ is the number of particles, or the number of proposal samples that will be weighted and re-sampled via the likelihood $p({\bf x}_M | {\bf z}_p(t))$. In practice, this method is implemented by maintaining $Q$ sets of token logits ("particles") during inference, and at each step 1) drawing a proposal sample from $p({\bf z}(t))$ independently for each particle and 2) re-sampling particles via $p({\bf x}_M | {\bf z}_p(t))$, replacing the previous particles with the re-sampled set. Since drawing proposal samples involves making a forward pass with the network $\hat{e}$ and taking a single step with the BFN SDE solver for each particle, this method incurs significant runtime costs as $Q$ increases. While ProtBFN's sampler used a value of $Q = 1024$, we found that competitive results could be achieved in IgCraft on a much smaller compute budget and used $Q = 32$ for all experiments. 

To allow for the specification of fixed sequence lengths during inpainting, IgCraft extends the likelihood function in \eqref{particle_weights} by subtracting the model log-likelihood assigned to pad tokens in the masked region. Using $\hat{e}_{\text{pad}}({\bf z}_p(t), t)_i$ to denote the probability assigned to a pad value at the $i^{\text{th}}$ sequence token, we use the following unnormalized particle weights:

\begin{equation}
   w_p = \sum_{i \in M}{
    -\lVert 
        \hat{e}({\bf z}_p(t), t)_i - e({\bf x}_i)
    \rVert^2
    } - 
    \sum_{j \notin M}{
    \log{
        \left [
        \hat{e}_{\text{pad}}({\bf z}_p(t), t)_j
        \right ]
    }
    }
    \label{fixed_length_particle_weights}
\end{equation}

In our experiments for sequence inpainting, inverse folding, and CDR grafting, we use the fixed-length likelihood function in \eqref{fixed_length_particle_weights} and expose pad tokens within the region(s) being inpainted via the mask $M$. 

\subsubsection{Gated residual connections}

A key element of IgCraft's network architecture is the use of gated residual connections \citep{Srivastava2015, Dhayalkar2024} at specific positions between layers that are updated at different stages of the training process (\autoref{figure:architecture}). A standard sigmoid-gated linear unit (GLU) \citep{Dauphin2016} is defined as follows:

\begin{equation}
    \text{GLU}({\bf x}) = \text{sigmoid}({\bf W}^{(g)}{\bf x} + {\bf b}^{(g)}) \odot ({\bf W}^{(l)}{\bf x} + {\bf b}^{(l)})
    \label{glu}
\end{equation}

Where ${\bf W}^{(g)}$, ${\bf b}^{(g)}$, ${\bf W}^{(l)}$, and ${\bf b}^{(l)}$ are all different parameters. A gated residual connection is then simply a residual connection between an input embedding ${\bf x}$ and a GLU-transformed hidden representation ${\bf h}$ output by another layer:

\begin{equation}
    \text{GatedResidual}({\bf x}, {\bf h}) = {\bf x} + \text{GLU}({\bf h}) 
\end{equation}

The key insight is that by initializing ${\bf b}^{(g)}$ to some large negative value (we use $-5.0$), we have $\text{GLU}({\bf h}) \approx {\bf 0}$ and $\text{GatedResidual}({\bf x}, {\bf h}) \approx {\bf x}$. In IgCraft, this allows the weights pre-trained on unpaired sequences (\autoref{figure:architecture}, blue) to serve as a suitable initialization for paired sequence fine-tuning, since initializing ${\bf b}^{(g)} = -5.0$ in the interaction blocks prevents any information from the other chain from entering each sub-network's residual stream, allowing the model to treat VH/VL pairs as two unpaired sequences at the start of paired sequence training. Likewise, the transformer backbone weights trained on paired sequences (\autoref{figure:architecture}, blue/green) serve as a starting point for paired structure fine-tuning, with the structure encoder's output being ignored at initialization. 

\subsection{Supplementary results and experimental details}

\subsubsection{Unconditional sampling}

Here we present additional metrics calculated for the unconditional sequences sampled from IgCraft and p-IgGen \citep{Turnbull2024} as well as the reference set of 2000 held-out test sequences. Germline statistics were calculated by performing alignment with ANARCI \citep{Dunbar2015}.

\begin{table}[h!]
\captionsetup{font=footnotesize}
\caption{Additional unconditional sampling metrics from each set of 2000 paired sequence samples. We report the mean germline sequence identity (\%) for V and J genes, the germline diversity (measured as the shannon entropy of the observed categorical distribution of V and J genes), and the humanness score from AbNatiV \citep{Ramon2024}. Germline alignment was performed using ANARCI \citep{Dunbar2015}. All metrics are presented for both heavy and light chains in the form VH / VL.}
\begin{center}
\resizebox{\textwidth}{!}{
\begin{tabular}{|c|c|c|c|c|c|}
\hline 
{\bf Samples} & {\bf V seq. id. (\%)} & {\bf \bf J seq. id. (\%)} & {\bf V diversity} & 
{\bf J diversity} & {\bf Humanness}
\\ \hline 
IgCraft & 95.8 / 97.5 & 95.4 / 95.0 & 3.92 / 3.58 & 1.86 / 2.03 & 92.8 / 98.7 \\
Reference & 95.8 / 97.5 & 95.6 / 95.4 & 3.91 / 3.59 &
1.86 / 2.07 & 92.9 / 98.8 \\
p-IgGen & 96.1 / 97.0 & 95.0 / 95.1 & 3.91 / 3.67 & 
1.91 / 2.09 & 93.0 / 98.7 \\
\hline
\end{tabular}
}
\label{table:unconditional_supp}
\end{center}
\end{table}

\subsubsection{Sequence inpainting}

Below we provide a more detailed view of the joint inpainting capabilities of IgCraft compared to competing masked language modelling approaches, showing AAR statistics for CDR and framework regions on the VH chain, the VL chain, and both chains. 

\begin{figure}[h!]
    \centering
    \includegraphics[scale=0.5]{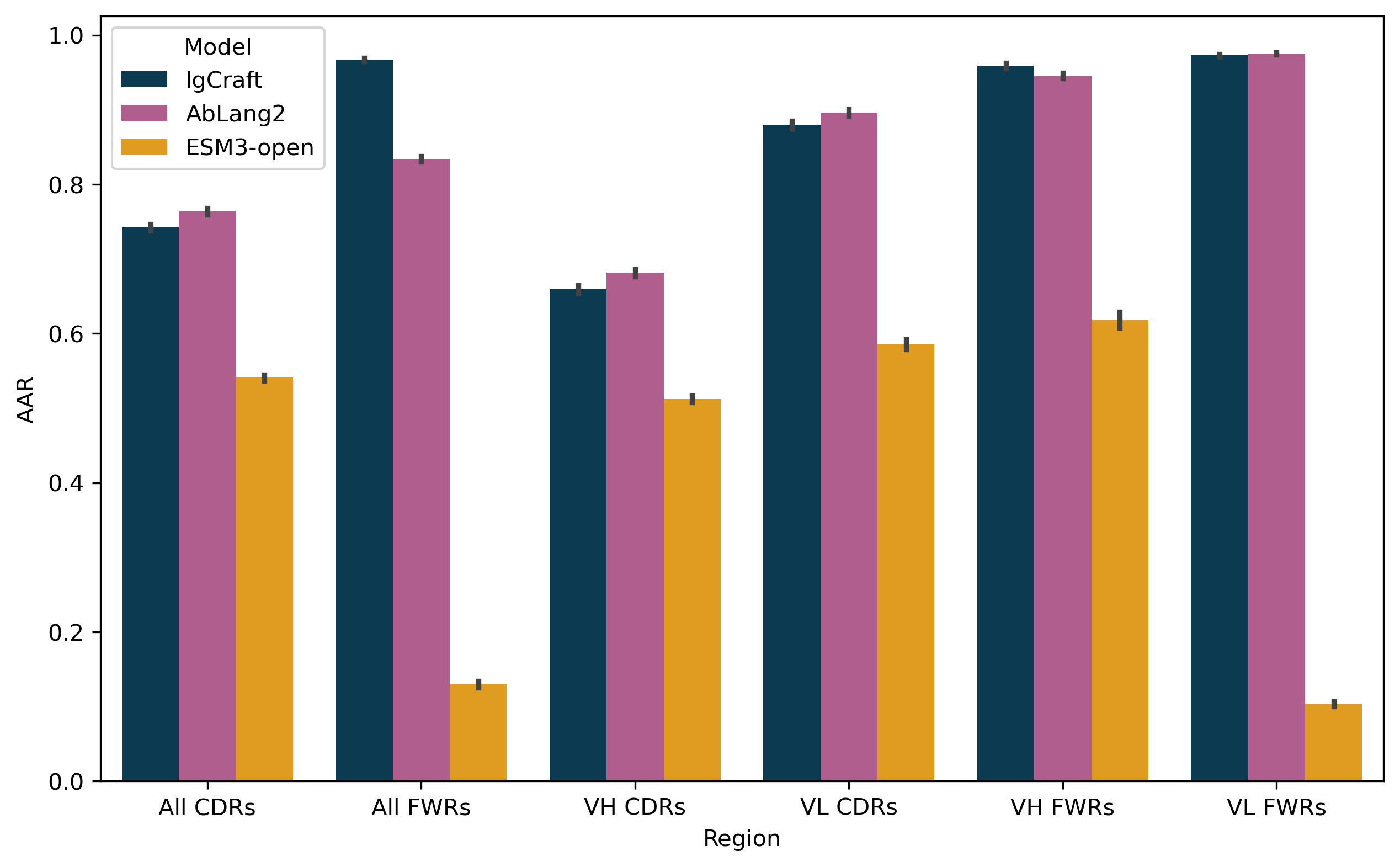}
    \captionsetup{font=footnotesize}
    \caption{Mean amino acid recovery (AAR) for sequence inpainting of different variable regions on the 2000 holdout sequences from paired OAS. Error bars correspond to the standard error of the estimated mean AAR.}
    \label{figure:joint_inpaint}
\end{figure}

\subsubsection{Structural data curation}

We implemented a stringent filtering process of SAbDab to ensure that our training and test datasets consist of non-redundant variable region structures and sequences for true paired VH/VL structures. Specifically, we use the following workflow for extracting structural data from each biological assembly PDB file in SAbDab:

\begin{enumerate}
    \item Extract the SEQRES field to obtain the true full-length sequence of each chain. If the SEQRES field is not present, extract the sequence from the ATOM records. Filter non-standard amino acids and heteroatoms.
    \item If a SEQRES field is present, perform a pairwise sequence alignment with scoring scheme match=1, mismatch=0, gap=0 between the sequence obtained from SEQRES and the sequence obtained from the ATOM records to obtain a mask corresponding to which amino acids appear in the ATOM records. This accounts for residues that could not be resolved in the structure, which can still be modelled with IgCraft via the sequence transformer backbone.
    \item Identify human heavy and light chains by performing sequence alignment with ANARCI using the IMGT numbering scheme. Discard antibody sequences which did not align to the human germline. Keep only the portion of each human antibody chain which was numbered. Keep all residues from non-antibody chains. 
    \item Identify heavy/light chain pairings by counting, for each heavy chain, the number of C$\alpha$-C$\alpha$ contacts (defined as \textless8\AA) with each light chain in the PDB file, and assume the light chain with the maximum number of contacts is the correct pairing. If no contacts are found with any light chain in the file, assume the chain is unpaired and discard it.
    \item Save each paired VH/VL sequence and perform a search for duplicates in the entire set of paired sequences extracted from SAbDab. If a chain pairing appears in multiple PDB IDs, take the entry with the lowest resolution (cryo-EM and NMR structures and labelled with resolution 0).
    \item For each remaining VH/VL pair, for non-antibody chains in the same file, retain the residue data from a maximum of 128 residues with the closest minimum C$\alpha$ distance to any residue in the antibody chains.
    \item From the remaining non-redundant, paired human antibody chains (with up to 128 target residues each), set aside as a test set all chains from PDB IDs which appear in the test set and not the training/validation sets of AbMPNN \citep{Dreyer2023}.
    \item From the remaining $\sim$2,700 structures, set aside 270 randomly-selected structures for monitoring validation loss during training. The remaining $\sim$2,400 structures are added to the training set of $\sim$30,000 predicted structures from paired OAS.
\end{enumerate}

\subsubsection{Inverse folding}

\begin{figure}[h!]
    \centering
    \includegraphics[scale=0.5]{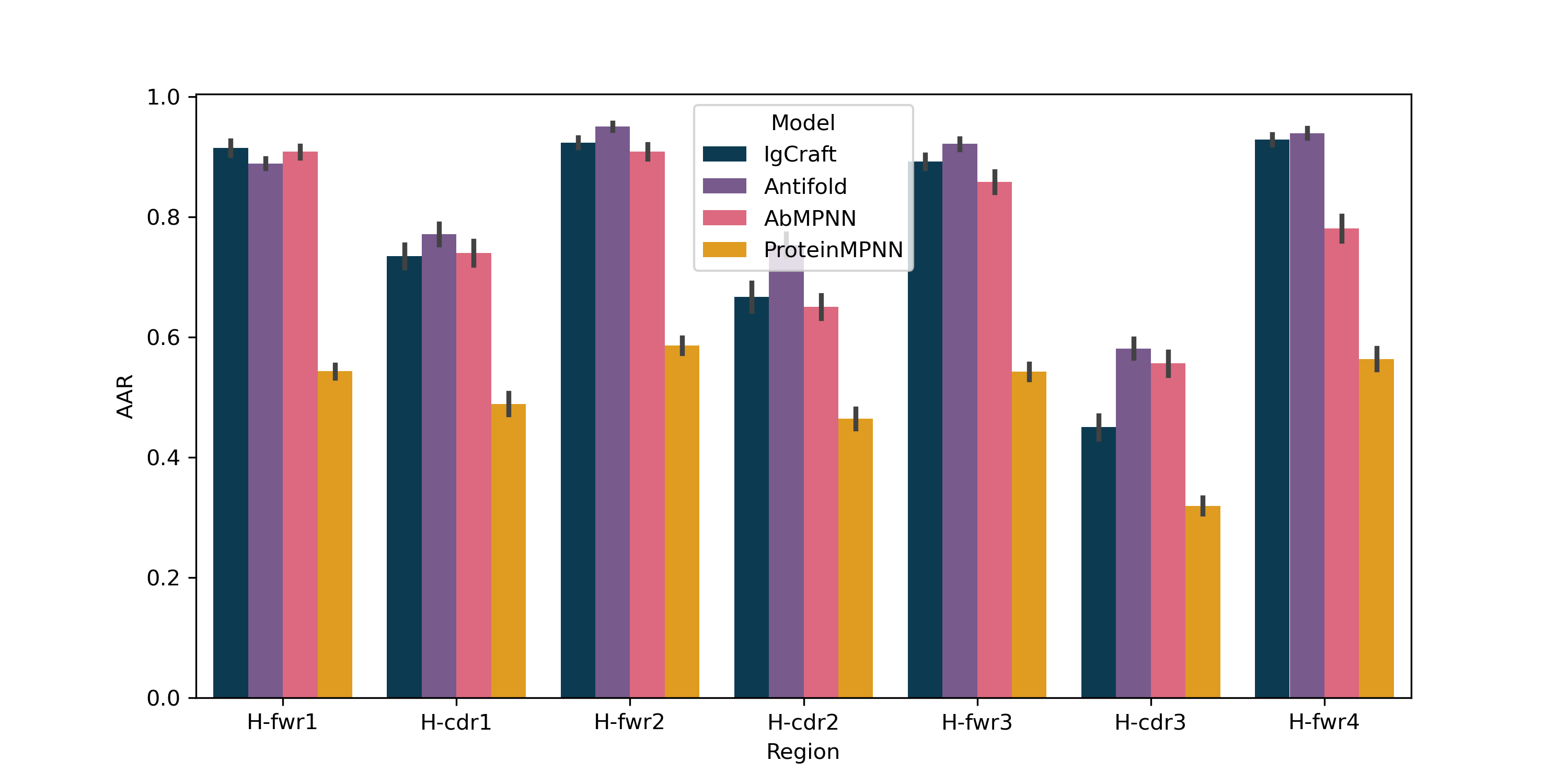}
    \captionsetup{font=footnotesize}
    \caption{Amino acid recovery (AAR) per-heavy chain region on 98 curated human antibody structures from the AbMPNN test set. Error bars correspond to the standard error of the estimated mean AAR.}
    \label{figure:vh_aar}
\end{figure}

\begin{figure}[h!]
    \centering
    \includegraphics[scale=0.5]{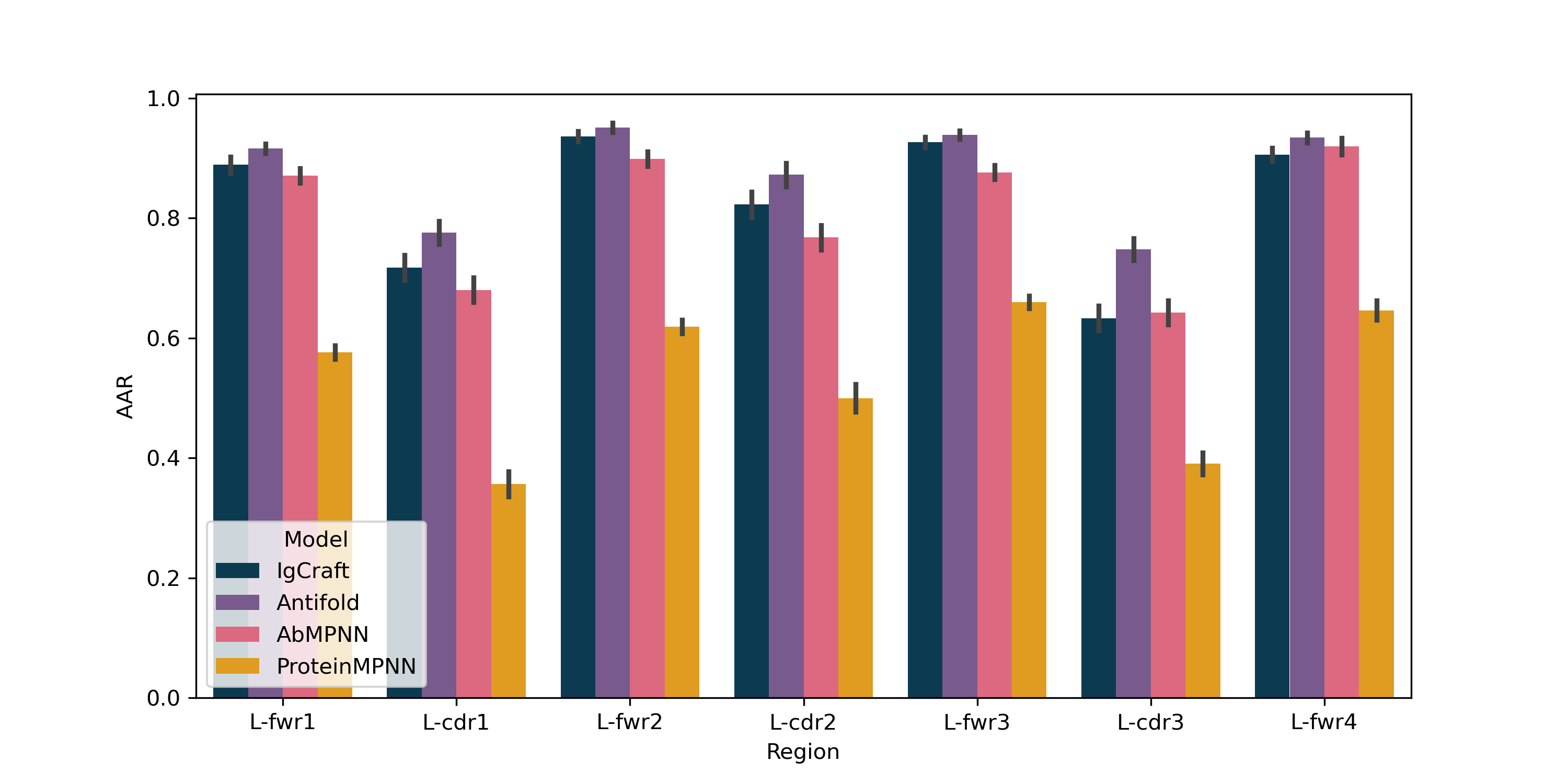}
    \captionsetup{font=footnotesize}
    \caption{Amino acid recovery (AAR) per-light chain variable region on 98 curated human antibody structures from the AbMPNN test set. Error bars correspond to the standard error of the estimated mean AAR.}
    \label{figure:vl_aar}
\end{figure}

Below we provide more detailed AAR statistics for each variable region with each model for the inverse folding task on 98 curated human antibody structures from the AbMPNN test set \citep{Dreyer2023}. We extracted from each source PDB file only the relevant paired antibody chains and any non-antibody chains with one or more C$\alpha$-C$\alpha$ contact (\textless8\AA) with any antibody residue. We then calculate amino acid recovery (AAR) per-region for each structure and report the mean AAR for each region over all structures (\autoref{figure:vh_aar} and \autoref{figure:vl_aar}).

\subsubsection{CDR grafting and humanisation}

To obtain data for CDR grafting we filtered for non-redundant paired mouse antibody structures from SAbDab (identified as mouse using germline alignment with ANARCI \citep{Dunbar2015}), using a cutoff date of February 2024. This yielded 27 unique paired mouse structures, 20 of which were bound structures containing non-antibody chains and 7 of which were unbound structures. We labelled CDRs for these antibodies according to IMGT definition. Using the structures of all CDRs as input to IgCraft, as well as the CDR sequences with a padding of two framework residues on each side, we sampled a single framework sequence for each set of mouse CDRs, using the structure encoder to encode the CDR structure and applying the particle filtering method in \eqref{fixed_length_particle_weights} with 32 particles to condition on the ground-truth (padded) CDR sequences. This produced full-length paired VH/VL sequences for each mouse antibody, containing the original mouse CDR sequences and generated framework sequences.

\begin{figure}[t]
    \centering
    \includegraphics[width=1.0\textwidth]{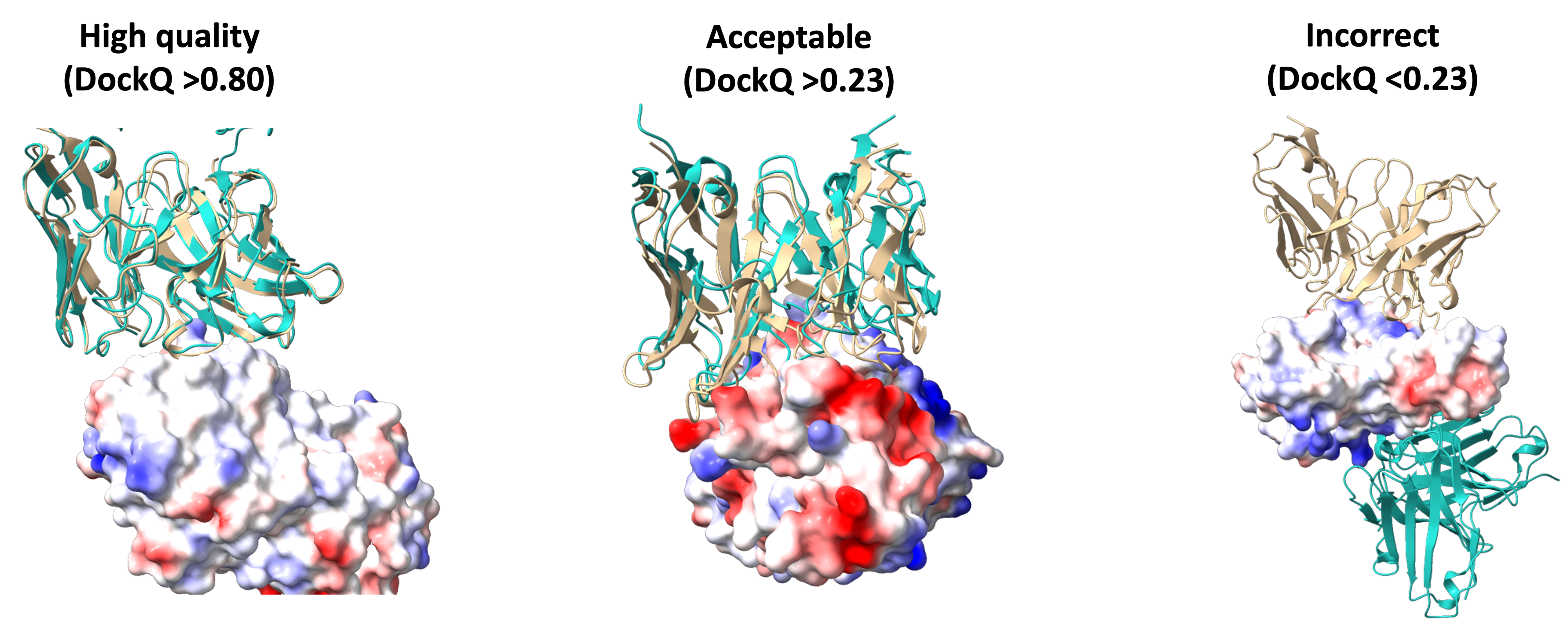}
    \captionsetup{font=footnotesize}
    \caption{Illustrative AlphaFold3 predictions for the grafted sequence and ground-truth bound crystal structures for the mouse antibodies used in the CDR grafting experiment. We show the ground-truth target protein as a surface with the predicted humanised VH/VL structure in gold and the ground-truth mouse antibody structure in blue. Three examples are chosen for visualization: the first (left) is a high-quality docked structure prediction with DockQ = 0.87 (PDB ID: 8TFH), the second (middle) is a docked structure with acceptable quality of DockQ = 0.26 (PDB ID: 8TXU), and the final (right) is an incorrectly docked structure with DockQ = 0.05 (PDB ID: 8TVH). We note that the WT mouse antibody for 8TVH (right) was also incorrectly docked by AF3, like most (9/10) of the grafted antibodies which produced DockQ \textless 0.23.}
    \label{figure:structures}
\end{figure}

To perform structure prediction with AlphaFold3 (AF3) \citep{Abramson2024}, we used ColabFold's multiple sequence alignment (MSA) pipeline \citep{Mirdita2022} to generate a paired MSA for each pair of antibody VH/VL chains, as well as a separate paired MSA for the target chains in PDB structures with more than one non-antibody chain. The paired-chain MSA is then split into segments (spans of columns) corresponding to the sequences of individual chains in each complex. We also generated an unpaired MSA for each chain in each file, and concatenated the unpaired MSA as new rows after each chain's portion of its paired MSA (if present). The final concatenated MSA was then used as an "unpaired MSA" input to AF3, where individual rows in the paired portion of MSAs for complexed chains correspond to sequences from the same species (unpaired target chains do not have a paired portion). We used these MSA inputs to run AF3 with 10 seeds and used the prediction with the highest ranking score as the final structural model (the default behavior). To calculate CDR RMSD statistics, we superimpose the framework regions of the corresponding antibody chain in the predicted and ground-truth structures, excluding CDR regions from being used to calculate the optimal superposition. We use the python implementation of DockQ to calculate docking scores \citep{Mirabello2024}. To provide some intuition as to what DockQ scores represent, we include three illustrative examples of AF3 structure predictions for grafted antibodies, including the ground-truth mouse antibody crystal structure but superimposing only using the target protein (\autoref{figure:structures}).

As an ablation study, we also performed CDR grafting using IgCraft without any structural information on the 27 test set mouse antibodies, otherwise using the same settings as in structure-conditioned generation (2 pad residues on both sides per-CDR, 32 particles). The results of this study are provided in \autoref{table:grafting_ablation}. Although IgCraft was able to achieve a high level of humanisation despite the lack of structural information, metrics obtained from the AF3 prediction (H-CDR3 RMSD, DockQ) indicate that giving IgCraft structural information significantly improved the model's ability to generate framework sequences that preserve the binding capability of the CDRs. We hope to explore in future work if using predicted structures in place of true crystal structures leads to a similar boost in performance in settings where ground-truth structural data is not available.

\begin{table}[t]
\captionsetup{font=footnotesize}
\caption{Structural ablation study for IgCraft in CDR grafting. We performed framework generation on 27
paired mouse antibody structures from SAbDab (20 bound, 7 unbound) using IgCraft without structure information (seq. only) and with both sequence and structure information (seq. + structure). We report the mean sequence identity (\%) between the grafted and original mouse sequences, the mean OASis humanness score \citep{Prihoda2022}, the mean AbNatiV humanness score per-chain \citep{Ramon2024}, the mean H-cdr3 C$\alpha$ RMSD between the AlphaFold3 predictions and each corresponding ground-truth PDB, and the fraction of the sequences for the 20 bound structures correctly docked by AF3 (DockQ \textgreater 0.23).}
\begin{center}
\resizebox{\textwidth}{!}{
\begin{tabular}{|c|c|c|c|c|c|}
\hline 
{\bf Samples} & {\bf \makecell{\% Seq. id. \\ (VH / VL)}} & {\bf \makecell{Humanness \\ (OASis)}} & {\bf \makecell{Humanness \\ (AbNatiV, VH / VL)}} & {\bf H-CDR3 RMSD (\AA)} & {\bf DockQ \textgreater0.23} 
\\ \hline 
Seq. only & 72.9 / 77.4 & 78.4 & 0.85 / 0.91 & 2.42 & 7/20 \\
Seq. + structure & 77.6 / 77.5 & 77.9 & 0.88 / 0.90 & 2.04 & 10/20 \\
\hline
\end{tabular}
}
\label{table:grafting_ablation}
\end{center}
\end{table}

\vfill

\end{document}